\documentclass[preprintnumbers,showpacs,twocolumn,amsmath,amssymb]{revtex4}

\usepackage{amsmath}
\usepackage{amsfonts}
\usepackage{amssymb}
\usepackage{verbatim}
\usepackage{graphicx}

\usepackage{color}
\usepackage{ulem} 

\definecolor{BLACK}{gray}{0}
\definecolor{WHITE}{gray}{1}
\definecolor{RED}{rgb}{1,0,0}
\definecolor{BLAH}{rgb}{.3,.3,.5}
\definecolor{GREEN}{rgb}{0.2,.6,0.2}
\definecolor{BLUE}{rgb}{0,0,1}
\definecolor{CYAN}{cmyk}{1,0,0,0}
\definecolor{MAGENTA}{cmyk}{0,1,0,0}
\definecolor{YELLOW}{cmyk}{0,0,1,0}
\definecolor{YELLOW2}{cmyk}{.2,.2,.8,0}

\definecolor{orange}{rgb}{0.9,0.3,0}

\newcommand{\tp}{^{\top}}

\begin{document}


\title{Loss-Tolerant Tests Of Einstein-Podolsky-Rosen-Steering}

\author{D.\ A.\ Evans,$^{1,2}$  E.\ G.\ Cavalcanti,$^{2,3,4}$ H.\ M.\ Wiseman$^{1,2}$}

\affiliation{$^{1}$Centre for Quantum Computation and Communication Technology (Australian Research Council), Griffith University, Brisbane, 4111, Australia\\ $^{2}$Centre for Quantum Dynamics, Griffith University, Brisbane, 4111, Australia\\ $^{3}$School of Physics, University of Sydney, NSW 2006, Australia\\
$^{4}$Quantum Group, Department of Computer Science, University of Oxford, United Kingdom}

\date{8 August 2013}

\begin{abstract}
We analyse two classes of Einstein-Podolsky-Rosen (EPR)-steering inequalities, the violation of which can be used to demonstrate EPR-steering with an entangled two-qubit Werner state: linear inequalities and quadratic inequalities. We discuss how post-selection of results (by appeal to the fair sampling assumption) can compromise the rigour of these inequalities in experimental tests of EPR-steering. By considering the worst-case scenarios in which detector inefficiency or other loss could be exploited within a local hidden-state model, we derive inequalities that enable rigorous but loss-tolerant demonstrations of EPR-steering. The linear inequalities, and special cases of the quadratic inequalities, have been used in recent experiments. Our results indicate that regardless of the number of settings used, quadratic inequalities are never better, and often worse, than linear inequalities.
\end{abstract}



\maketitle

\section{Introduction}
The nonlocality of entangled quantum states is one of the most significant, and most heavily debated, features of quantum mechanics. It was an apparent action-at-a-distance which was Einstein, Podolsky, and Rosen's basis for challenging the completeness of quantum mechanics in their historic paper~\cite{key-1}. There they demonstrated, starting from a set of assumptions now known as ``local realism" \cite{despag}  or ``local causality'' \cite{locaus}, 
a contradiction between the quantum predictions for certain correlations between entangled states and the assumption that quantum mechanics provides a complete description of reality.

Schr\"odinger \cite{Schrod} noticed the strangeness of the nonlocal effect uncovered by EPR, which he called ``steering,'' but believed that such correlations would not be observable between distant objects. In 1964 Bell \cite{Bell} showed that EPR's assumption of local causality was itself in contradiction with the predictions of quantum mechanics, and that no local hidden variable (LHV) model of the kind envisaged by EPR could explain all correlations between entangled states. Those correlations (i.e., the violation of Bell inequalities) have been since then observed \cite{Aspect,Giustina}, up to some experimental loopholes \cite{Garg,Cyril}.

Schr\"odinger \cite{Schrod} also coined the term ``entanglement," but the concept was not defined for general (mixed) states until 1989, by Werner \cite{Werner}. In that paper, Werner also showed that not all entangled states violate a Bell inequality, thus demonstrating that Bell nonlocality is not synonymous with entanglement \footnote{Werner actually used the term ``EPR correlated'' instead of ``entangled'', but as we'll see below, EPR correlations are not synonymous with entanglement either.}.

Inequalities to demonstrate the correlations of the EPR paradox have been proposed, also in 1989, by Reid and co-workers \cite{Reid1989,Reid2009}. However, the correlations discussed by EPR were formally defined as a general class only in 2007, by Wiseman and co-workers \cite{key-3,key-4}. Adopting Schr\"odinger's term for the EPR correlations, they showed that not all entangled states display steering, and not all states that demonstrate steering violate Bell inequalities. To avoid ambiguity in terminology, this effect is now usually labelled \textit{EPR-steering} \cite{key-5,Saunders,Bennett,Parsimonious,Vallone}. 

A formal definition of EPR-steering allows mathematically rigorous tests to be devised \cite{key-5}, which are typically formulated as inequalities. Experimental results which violate these inequalities should then demonstrate EPR-steering (much as Bell nonlocality is demonstrated by the violation of Bell inequalities). However, as is the case with Bell inequalities, there are gaps to be bridged between mathematical and experimental rigour. Use of the fair sampling assumption results in one such loophole through which experimental imperfections may compromise the rigour in tests of EPR-steering \cite{Garg,Bennett}. 

In this paper, we will describe two different kinds of EPR-steering criteria -- linear and nonlinear -- which address the inefficient detection loophole, each with a varying number of qubit spin measurements (regularly spaced about the Bloch sphere), from 2 to 10. From these criteria, we calculate loss-tolerant EPR-steering inequalities that close the detection loophole, and we then compare the strength of the various tests. The linear inequalities that we derive are the same as those used in Ref.~\cite{Bennett}, and special cases of the nonlinear inequalities that we derive were used in Refs.~\cite{Vienna,UQ}. We derive lower bounds on the efficiency of detectors that can be used to rigorously demonstrate EPR-steering. It will be observed that in the great majority of cases, using more measurements makes our tests of EPR-steering more loss-tolerant. It will also be observed that the nonlinear inequalities we consider offer no advantage over the linear inequalities. All of our inequalities are optimized for Werner states (that is, depolarized versions of maximally entangled states); loss-tolerant EPR-steering inequalities for non-maximally entangled states are considered in Ref.~\cite{Vallone}. 

The remainder of this paper is organised as follows. We begin Sec.~II by describing EPR-steering, and its operational definition. We introduce two different EPR-steering criteria in Sec.~III, and discuss Bob's optimal measurement strategies for using these criteria. We then address, in Sec.~IV, the ways in which Alice and Bob can deal with the possibility of inefficient detectors. Following from this discussion, we derive loss-tolerant bounds on both EPR-steering inequalities in Sec.~V. Finally, we compare how well these criteria perform for demonstrations of EPR-steering.

\section{Definition of EPR-steering}

A demonstration of EPR-steering is a task involving two parties, Alice and Bob, who share a pair of quantum systems that may or may not be entangled (in this paper, we restrict our considerations to qubits). The task is for Alice to convince Bob that they share entanglement. However, while Bob trusts his own apparatus in that he trusts that his measurements are described by the appropriate quantum mechanical operator on a qubit, he does not trust Alice (or her apparatus) in the same way.

The task proceeds as follows: Upon receiving a state, Bob tells Alice which measurement to make, and performs the same measurement on his own state (this is optimal for verifying singlet-like entanglement, which is typically the case of interest \cite{key-3,key-4,key-5,Saunders,Parsimonious,Vallone,Bennett,Vienna,UQ,preparation}). We assume that Bob's equipment is trustworthy, and therefore is actually performing measurements on a quantum state to generate his results. However, Bob does not trust that 
Alice is 
generating her results by performing measurements on a state that is entangled with his, but rather allows for the possibility that
his state could be locally predetermined by some variables which Alice may have access to. This would amount to a local hidden state \cite{key-4} (LHS) model for Bob's system, where Bob's results are derived from some quantum state, but where no such assumption is made about Alice's results, which may be determined by some hidden variable (which may be classically correlated with Bob's state). Thus, if the experimental statistics cannot be described by a LHS model, then Bob will be convinced that Alice can steer his state, and thus that they share entanglement. As shown in Ref.~\cite{key-5}, the assumption of a LHS model leads to certain \textit{EPR-steering inequalities}, the violation of which indicate the occurrence of EPR-steering.

Without being able to nonlocally influence Bob's state (i.e., in a LHS model), Alice's ability to manipulate Bob's measurements is maximised if she can control what pure state Bob will be given. Thus, we will assume that she has the ability to do so. In the operational scheme we are considering, Alice knows which measurements Bob will make, but she does not know which one he will choose in any given run of the experiment. The causal separation of Alice and Bob is embodied in the assertion that Bob selects his measurements in such a way that Alice must decide what state to send Bob before she knows what measurement he will perform on it.

\section{EPR-Steering Inequalities}

The two kinds of EPR-steering criteria in this paper are both additive convex criteria, employed in the manner described by Ref.~\cite{key-5}. The first will be a linear correlation function between Bob's measured results and Alice's claimed results. The second will be based on Bob's \textit{inference variance} \cite{Reid1989}, i.e., the conditional variance of Bob's measurement results, given Alice's.

The entangled states that we will consider are Werner states, i.e. states of form
\begin{equation}
\rho^{\alpha\beta}=\mu|\psi_s \rangle\langle\psi_s|+(1-\mu)\frac{\mathbb{I}^{\alpha\beta}}{4},\label{eq:werner}
\end{equation}
where $|\psi_s\rangle$ is the spin singlet state $|\psi_s\rangle=\frac{1}{\sqrt{2}}\left(|0\rangle^{\alpha}\otimes|1\rangle^{\beta}-|1\rangle^{\alpha}\otimes|0\rangle^{\beta}\right)$, and the purity parameter $\mu$ is constrained such that $0\leq\mu\leq1$. The superscript $\alpha$ denotes a feature of Alice's system, and $\beta$ denotes a feature of Bob's. The experiments to which our work applies \cite{Saunders,Bennett,Vienna,UQ} were done using  entangled pairs of photon polarisation qubits, but we will use the terminology of spin.

\subsection{Additive correlation bound}

The first EPR-steering criterion that we will use is the average (over the measurement settings) of the expectation value of the correlation function of Alice and Bob's spin measurements, which we will label $S_{n}$. It can be defined generally (i.e., without assuming an honest Alice) from Alice's expectation value of Bob's result, $\langle \hat{\sigma}_{j}^{\beta}\rangle$, based on her own result, $A_j$:
\begin{equation}
S_{n}=-\frac{1}{n}\sum_{j=1}^{n} E_{A_j}\left[  A_{j} \langle \hat{\sigma}_{j}^{\beta}\rangle_{A_j} \right].
\label{eq: Sn}
\end{equation}
Here the index $j$ references a measurement setting from a set $\{j\}$ of $n$ different orientations along each of which Alice and Bob must measure. Bob must randomly choose between these $n$ measurements, and since they are in separate labs, Alice does not know which measurement will be performed on each state she sends to Bob. It is this feature which will lead to the limits on $S_{n}$ in a no-steering model. The symbol $E_{A_j}$ denotes the ensemble average over Alice's result $A_j \in \{-1,1\}$.

For an honest Alice (in a quantum mechanical model), Eq.~(\ref{eq: Sn}) can be expressed as
\begin{equation}
S_{n}=-\frac{1}{n}\sum_{j=1}^{n}\langle\hat{\sigma}_{j}^{\alpha}\hat{\sigma}_{j}^{\beta}\rangle
= -\frac{1}{n}\sum_{j=1}^{n} \sum_{A_j} P(A_j)  A_{j} \langle \hat{\sigma}_{j}^{\beta}\rangle_{\rho_{A_{j}}^{\beta}},
\label{eq: honest}
\end{equation}
where $\rho_{A_{j}}^{\beta}$ is the reduced state of Bob's qubit, conditioned on outcome $A_j$ for Alice's measurement of $\hat{\sigma}_{j}^{\alpha}$. Correspondingly, $B_{j}$ shall represent the result of Bob's measurement of  $\hat{\sigma}_{j}^{\beta}$, when necessary. In a quantum mechanical model where Alice is honest, and Alice and Bob do share a Werner state as in Eq.~(\ref{eq:werner}), the correlation function yields a result 
\begin{equation}
S_{n}(\rm{Werner})=\mu.
\label{WernerMu}
\end{equation}
It should be noted that for individual measurements of a Pauli operator, the results of either party's measurements will always be either $1$, or $-1$. However, if Alice is cheating, there is no such constraint on what her measurements can be. If Alice is simply making up her results, she can choose any value she fancies, 
but if she submits to Bob any measurement other than $\pm 1$, he will recognise that her result was clearly not obtained from any spin measurement. Therefore, if an untrustworthy Alice is to convince Bob of EPR-steering, she must still constrain her results to $A_{j} = \pm 1$.

For an untrustworthy Alice who does not share entanglement with Bob, the experimental results must be describable by a LHS model and thus take the form
\begin{equation}
S_{n}=-\frac{1}{n}\sum_\xi  P(\xi)\sum_{j=1}^{n} A_{j,\xi} \langle \hat{\sigma}_{j}^{\beta}\rangle_{\rho_{\xi}^{\beta}},\label{eq: dishonest}
\end{equation}
where $\rho_{\xi}^{\beta}$ is the state which Alice prepares for Bob, with probability $P(\xi)$, and where $A_{j,\xi}$ is the result that Alice will declare for each value of $j$ and $\xi$. The choice of which state to send to Bob and the choice of $A_{j,\xi}$ can be said to have a common dependence on some variable, or set of variables, which we label $\xi$.  Alice's cheating strategy will thus consist of specifying these variables and dependencies so as to maximise $S_{n}$.

The assumption of a local hidden state model in Eq.~(\ref{eq: dishonest}) means that there exists a bound upon Eq.~(\ref{eq: dishonest}) that is not present upon Eq.~(\ref{eq: honest}). The general proof for linear EPR-steering criteria can be found in Ref.~\cite{key-5} and does not need to be reproduced here -- the relevant mathematical implication is that the
(achievable)
upper bound on the expectation value of an operator is the largest eigenvalue of that operator. Using this property, we can calculate the largest possible value of $S_{n}$ that can be obtained in a no-steering model. This means that for this case, where Alice reports a result in every run, Alice can maximize the correlation function $S_n$ by sending Bob an identical LHS for every run in an experiment. (The non-deterministic cases where she must choose different states will be dealt with in Sec.~IV.) Thus Alice's optimal choice of $\xi$ will be such that $\rho_{\xi}^{\beta}$ makes $\sum_j | \langle \hat{\sigma}_{j}^{\beta}\rangle_{\rho_{\xi}^{\beta}}|$ as large as possible. To obtain the maximum value of $S_n$ with this $\xi$, Alice must also use an optimal set of $A_{j,\xi}$ values. Such a set is easily found by requiring $A_{j,\xi}=1$ when $\langle \hat{\sigma}_{j}^{\beta}\rangle_{\rho_{\xi}^{\beta}} <0$ and $A_{j,\xi}=-1$ when $\langle \hat{\sigma}_{j}^{\beta}\rangle_{\rho_{\xi}^{\beta}} >0$. Regardless of the $\{A_{j,\xi}\}$ values that a cheating Alice may submit, it will hold that
\begin{equation}
-\frac{1}{n}\sum_{j}A_{j,\xi}\langle\hat{\sigma}_{j}^{\beta}\rangle_{\rho_{\xi}^{\beta}}
\ \leq \ \max_{\{A_{j}\}}  \left[\lambda_{{\rm max}}\left(-\frac{1}{n}\sum_{j}A_{j}\hat{\sigma}_{j}^{\beta}\right)\right].
\end{equation}
where $\lambda_{{\rm max}}$ denotes the maximum eigenvalue of the following operator. Note that we can drop the $\xi$ dependence of $A_{j,\xi}$ when performing the maximisation, and in future we will often use $A_{j}$ to denote the random variables that may actually depend upon $\xi$. Thus we have derived an EPR-steering inequality
\begin{equation}
S_n \leq k_{n} \equiv\  \underset{\{A_{j}\}}{{\rm max}}\left[\lambda_{{\rm max}}\left(\frac{1}{n}\sum_{j}A_{j}\hat{\sigma}_j^{\beta}\right)\right],\label{eq:kbound}
\end{equation}
the violation of which would demonstrate EPR-steering. The value of $k_{n}$ thus depends only on the set of measurements $\{\hat{\sigma}_j^\beta\}$. We will specify below that each set of $n$ measurements is uniquely characterised (among our five different measurement sets) by the number of measurements in it, so it is sufficiently distinctive to write that $k_{n}$ depends only on $n$. Note that in Eq.~(\ref{eq:kbound}) we have ignored the minus sign in the maximisation of eigenvalues as the eigenvalues of every possible $A_{j}\hat{\sigma}_j^{\beta}$ will come in pairs of real numbers that are of the same magnitude, but opposite sign.

\subsection{Additive inference variance bound }

Our second EPR-steering criterion is based on the variance of Bob's measurements,
\[
V_{j}=\langle\hat{\sigma}_{j}^{2}\rangle-\langle\hat{\sigma}_{j}\rangle^{2}=1-\langle\hat{\sigma}_{j}\rangle^{2}.
\]
The sum of Bob's variances over all $n$ measurements would be
\[
\sum_{j}V_{j}=\sum_{j}1-\langle\hat{\sigma}_{j}^{\beta}\rangle^{2}=n-\sum_{j}\langle\hat{\sigma}_{j}^{\beta}\rangle^{2},
\]
so it would be more notationally convenient to simply use a form of the second term in this expression. Labelling it as $W_{n}$ and including a normalisation factor of $-\frac{1}{n}$, we will concern ourselves with
\begin{equation}
W_{n}=\frac{1}{n}\sum_{j}^{n}\langle\hat{\sigma}_{j}^{\beta}\rangle^{2}
\end{equation}
which is another expression that satisfies the additive convex criteria described in Ref.~\cite{key-5}. (It should be noted that this function is strictly convex because all $\langle\hat{\sigma}_{j}^{\beta}\rangle$ values are necessarily real numbers -- if they were not, then $W_n$ would not necessarily be a convex function.) The way in which this can be used as an EPR-steering criteria is if we consider how well Alice can estimate the expectation value of $\hat{\sigma}_{j}^{\beta}$, conditioned upon her measurement result, $A_j$. Thus, our EPR-steering function will be 
\begin{equation}
W_{n}=\frac{1}{n}\sum_{j}^{n}E_{A_j}\left[\langle\hat{\sigma}_{j}^{\beta}\rangle^{2}_{A_j}\right],
\label{eq:wn}
\end{equation}
where, as above, $E_{A_j}$ represents Alice's ensemble average. A sufficiently large value for this parameter means Alice has a sufficiently small inference variance for a noncommuting set of Bob's observable that this knowledge can only be explained by the EPR-steering phenomenon. 

Estimates of Bob's results can be inferred from Alice's results. For an honest Alice, the accuracy of such inferred results would be dependent on the degree of entanglement present. In the case of a completely entangled state (i.e., $\mu = 1$), Bob's results could be determined from Alice's with perfect accuracy.


If Alice and Bob genuinely do share a Werner state as in Eq.~(\ref{eq:werner}), and Alice performs upon it a measurement in a direction $-{\bf r}$, then the conditioned state on Bob's side will be $\rho^{\beta}=(1-\mu)\mathbb{I}+\mu\rho_{\bf r}$. 
Here, $\rho_{\bf r}$ is a pure state with unit-length Bloch vector ${\bf r}$, representing the spin orientation (see also Sec.~III~C below).
The expectation value of a measurement $\hat{\sigma}_{j}$ on a pure state $\rho_{\bf r}$ is $\langle\hat{\sigma}_{j}\rangle_{\rho_{\bf r}}={\rm Tr}[\hat{\sigma}_{j}\rho_{ \bf r}]=\mathbf{b}_j\cdot\mathbf{r}$, where $\mathbf{b}_j$ is the vector representing the orientation of the measurement $\hat{\sigma}_{j}$.
In the general case, Bob's expectation value $\langle\hat{\sigma}_{j}\rangle_{\rho^\beta}$ is given by
\begin{eqnarray*}
\langle\hat{\sigma}_{j}\rangle_{\rho^{\beta}} & = & {\rm Tr}[\hat{\sigma}_{j}(1-\mu)\mathbb{I}]+{\rm Tr}[\hat{\sigma}_{j}\mu\rho_{\bf r}]\\
 & = & \mu{\rm Tr}[\hat{\sigma}_{j}\rho_{\bf r}]=\mu\mathbf{b}_j\cdot\mathbf{r},
\end{eqnarray*}
which obviously gives $\langle\hat{\sigma}_{j}\rangle_{\rho^{\beta}}=\mu$ when $\mathbf{b}_j=\mathbf{r}$. In a quantum mechanical model
, Eq. (\ref{eq:wn}) can be calculated as 
\begin{equation}
W_{n}=\frac{1}{n}\sum_{j}^{n}\sum_{A_{j}}P(A_{j})\langle\hat{\sigma}_{j}^{\beta}\rangle_{\rho_{A_{j}}^{\beta}}^{2}.
\label{eq:honestWn}
\end{equation}
Thus, when an honest Alice is directed by Bob to measure in directions aligned (or anti-aligned) to his own, we should obtain, for all $n$-values,
\begin{equation}
W_{n}=\mu^{2}.
\end{equation}
%
On the other hand, if Alice is dishonest, and does not share entanglement with Bob, her results will be determined by a LHS model, and $W_n$ will be calculated as
\begin{equation}
W_{n}=\frac{1}{n}\sum_{\xi}P(\xi)\sum_{j}^{n}\langle\hat{\sigma}_{j}^{\beta}\rangle_{\rho_{\xi}^{\beta}}^{2}.
\label{eq:dishonestWn}
\end{equation}
In the absence of any entanglement, the accuracy with which Alice can infer Bob's results is dependent on how well Alice can predict the results of Bob's measurements upon his LHS, $\rho_{\xi}^{\beta}$. Therefore, Alice's optimal choice of LHS for Bob will be a state which minimises the summed variance of Bob's results (thus maximising $\sum_j  \langle \hat{\sigma}_{j}^{\beta}\rangle_{\rho_{\xi}^{\beta}}^2$).

Just as with $S_n$, the assumption of a LHS model in Eq. (\ref{eq:dishonestWn}) means that there is a bound upon Eq. (\ref{eq:dishonestWn}) that does not exist for Eq. (\ref{eq:honestWn}).
The bound on this function can be derived from the bound on the average expectation value of $\hat{\sigma}_{j}^{\beta}$. 
%
If Alice sends Bob a pure state $\rho_{\bf r}$, we will have
\begin{eqnarray}
\sum_{j}\langle\hat{\sigma}_{j}\rangle^{2}_{\rho_{\bf r}} & = & \sum_{j}(\mathbf{b}_j\cdot\mathbf{r})^{2}=\sum_{j}\mathbf{b}_j\cdot\mathbf{r}\mathbf{r}\cdot\mathbf{b}_j\nonumber\\
 & = & \sum_{j}\mathbf{b}_j^{T}\mathbf{r}\mathbf{r}^{T}\mathbf{b}_j=\mathbf{r}^{T}\left(\sum_{j}\mathbf{b}_j\mathbf{b}_j^{T}\right)\mathbf{r}\nonumber\\
 & = & \mathbf{r}^{T}\mathbf{N}\mathbf{r}\leq\lambda_{{\rm max}}(\mathbf{N}),
\label{gnworking}
\end{eqnarray}
where we have labelled the matrix $\sum_{j}\mathbf{b}_j\mathbf{b}_j^{T}$ as $\mathbf{N}$, and have used the fact that $|\mathbf{r}|=1$ for pure states. Using this inequality, the bound on $W_{n}$ in a no-steering (LHS) model can be seen to be 
\begin{equation} \label{AVEPRS}
W_{n}\leq g_n \equiv\ \frac{1}{n}\lambda_{{\rm max}}(\mathbf{N}).
\end{equation}

This bound is derived for the case that a dishonest Alice uses only one state, $\rho_{\bf r}$, to repeatedly send to Bob, but it is clear that it still holds for the case of Alice sending multiple states (so long as Bob always chooses his measurements in an order that is unpredictable by Alice) by the convexity of the function $W_n$ \cite{key-5}. That is, if we consider a mixed state, like $\rho=P_{1}\rho_{r_{1}}+P_{2}\rho_{r_{2}}$, $\langle\hat{\sigma}_{j}\rangle_{\rho}^2 = (P_{1}\mathbf{b}_j\cdot\mathbf{r}_{1}+P_{2}\mathbf{b}_j\cdot\mathbf{r}_{2})^2 \leq P_1 (\mathbf{b}_j\cdot\mathbf{r}_{1})^2 + P_{2}(\mathbf{b}_j\cdot\mathbf{r}_{2})^2$, where $P_{1}$ and $P_{2}$ are probabilities such that $P_{1}+P_{2}=1$.
Thus Eq.~(\ref{AVEPRS}) is an EPR-steering inequality.

\subsection{Bob's measurement scheme}

In choosing a measurement scheme, one must consider the ways in which measurement arrangements could be exploited by a cheating Alice. The bounds that we will derive are bounds on the average values of certain results, and thus, to experimentally determine these averages, the same measurements must be made many times, by both parties. So it is fair to say that Bob cannot realistically keep Alice from knowing the set of measurements he is to make. However, it is possible for Bob to prevent Alice from predicting the ordering of each measurement by randomising his choice of measurement in each run of the experiment \footnote{This assumes that Alice does not have access to Bob's random number generator, which is addressed in other works as the Free Will assumption \cite{Bell}.}.

For the EPR-steering functions introduced above we can assume that Alice sends Bob the same state in every run. This is because if Alice does not know the order in which Bob makes his measurements, each state she uses is equally likely to face any one of Bob's measurements. Therefore, her best option is to use the one state that has the highest average correlation function over all of Bob's measurements. If there exists more than one state which fulfils this requirement, then Alice may just as easily use a combination of them, but cannot possibly attain a maximum correlation function any higher than is attainable by using just one such state. Thus, in obtaining the maximal bound $k_{n}$ or $g_n$, there is never any advantage for Alice to use more than one state per experiment. 

Conversely, it is a good strategy for Bob to choose as many different measurements as possible (a large $n$ value). Indeed, if Bob chooses just one measurement orientation, then Alice will be able to align the spin axis of his state with that measurement axis every time. In that case, Alice could perfectly imitate the results of entanglement, obtaining $S_{n=1}=1$ or, equally, $W_{n=1}=1$.

For similar reasons, it is clear that if Bob chooses measurements that are close to each other on the Bloch sphere, it will lead to a higher $k_{n}$ or $g_n$ than if he chose measurements farther away from each other. 
This is bad for an honest Alice, as it means she would have to use a Werner state with higher purity $\mu$ to beat that bound, which is not what Bob wants.
Therefore it would seem likely that a good choice for Bob would be to make his measurements regularly spaced on the Bloch sphere. For this reason, we will follow Refs.~\cite{Saunders,Bennett} in choosing sets of measurement orientations that are related to the vertices of the three-dimensional Platonic solids. 

There are only five three-dimensional Platonic solids: those with 4, 6, 8, 12, and 20 vertices. It is important to note that every Platonic solid except that with four vertices (the tetrahedron) has vertices such that every vertex has a diametric opposite -- an antipode -- as some other vertex on the solid. This property is essential if we are to associate vertices with qubit observables, as every operator  $\hat{\sigma}_{j}$ has two eigenvectors which are antipodal on the Bloch scheme. Thus we will characterise each set of measurements by the number of measurement axes, which will be half of the number of vertices -- and will be denoted $n$. However, since the tetrahedron possesses four vertices, none of which are antipodes, we cannot use it to define an $n=2$ set of measurement axes (we could use the four different directions to define an $n=4$ set, but that would duplicate the cube). But we can still use another shape for $n=2$: the square -- a shape with four vertices, each of which is an antipode of one other vertex, and which are all regularly spaced, albeit in two dimensions. Thus, our measurement schemes involve Bob's chosen sets of measurements $\{j\}$ being of size $n=2$, 3, 4, 6, or 10.

As referenced above, these Platonic solid measurement schemes have been previously employed, with the bound $k_n$ having been calculated in those publications \cite{Saunders,Bennett} for $n=2$, 3, 4, 6, and 10. In those papers, it was calculated that (when Alice reports results with perfect efficiency) the value of $k_n$ was a monotonically decreasing function of $n$ (and strictly decreasing for every point except for $k_3 = k_4$). Specifically, it was found that $k_2 = 1/\sqrt{2} \approxeq 0.707$, $k_3 = k_4 = 1/\sqrt{3} \approxeq 0.577$, $k_6 \approxeq 0.539$, and $k_{10} \approxeq 0.524$.

However, the additive inference variance bound, $g_n$, has not been previously calculated for this scheme. In calculating the bound $g_n$ from Eq.~(\ref{AVEPRS}), we find that $g_2=\frac{1}{2}$ and $g_3=\frac{1}{3}$, but also that $g_3=g_4=g_6=g_{10}$. Although $g_3 < g_2$ as with $k_n$, $g_n$ does not decrease with $n$ beyond $n=3$. This happens because each measurement set for $n\geq 3$ forms a spherical 2-design \cite{Hong:1982}, which, for the form of $W_n$, means that each $g_n$ bound for $n \geq 3$ requires an equally high $\mu$-value to violate.

A simplified explanation for this can be drawn from the observation that for $n=2$, we use measurement vectors $\mathbf{b}_x=\left(1, 0 \right)\tp$,  $\mathbf{b}_y=\left(0, 1 \right)\tp$, and thus $\mathbf{N}=\mathbb{I}_2$. The same behaviour, but in three dimensions, occurs for the $n=3$ measurement vectors. The same pattern obviously does not hold for the rest of the measurement sets, since they are all in three spatial dimensions, and none of the other $\mathbf{N}$ matrices are equal to the identity, but the rest of the measurement sets are spherical 2-designs, meaning that for every other $n$-value, $\mathbf{N}\propto \mathbb{I}_3$. For this reason, each of these measurement sets exhibit the same behaviour in a test of EPR-steering. However, this is not the case for $\mathbf{N}$ matrices constructed from only a portion of the values in some $\{\mathbf{b}_j\}$ set, a fact which will become significant in Sec.~V. 

\subsection{Alice's cheating strategies} 
Having considered Bob's measurement strategies, we must also consider the optimal ways in which a dishonest Alice can attempt to exploit them. It is from just such an analysis that we can determine whether the EPR-steering bounds derived above are tight. That is, we will determine whether Alice can saturate the bound by sending Bob a state (i.e. a LHS) drawn from some ensemble $\{\rho_{\xi}^{\beta}\}$ over $\xi$.

Intuitively, the heart of Alice's cheating strategies in the situations we consider is to orient Bob's state as closely as possible to as many of his measurements as possible. This is because for maximising $S_n$ or $W_n$, Alice needs Bob's expectation values to be as large as possible.

This is easily shown for $S_n$, whose maximal bounds are calculated from $\lambda_{\max}\left(\frac{1}{n}\sum_{j}A_{j}\hat{\sigma}_{j}^{\beta}\right)$, into which we shall substitute $\hat{\sigma}_{j}^{\beta}=\mathbf{b}_{j}\cdot\boldsymbol{\sigma}$, where $\mathbf{b}_{j}$ is the vector orientation of $\hat{\sigma}_{j}^{\beta}$, with $\boldsymbol{\sigma}=(\sigma_{x},\sigma_{y},\sigma_{z})$. We will use the relation that for a general operator $\hat{O}$ on a state $\rho$, $\lambda_{\max}(\hat{O})=\underset{\rho}{\max}(\rho\hat{O})$, with the maximum being obtained by a pure state. Defining the state of Bob's system as $\rho_{\bf r}$, a pure state aligned with orientation $\mathbf{r}$, we find that
\begin{eqnarray}
\max(S_{n})|_{\{A_j\}}  & = &  \lambda_{\max}\left(\frac{1}{n}\sum_{j}A_{j}\mathbf{b}_{j}\cdot\boldsymbol{\sigma}\right)\nonumber\\
& = & \underset{\rho_{\bf r}}{\max}\,{\rm Tr}\left[\rho_{\bf r}\frac{1}{n}\sum_{j}A_{j}\mathbf{b}_{j}\cdot\boldsymbol{\sigma}\right]\nonumber\\
 & = & \underset{\mathbf{r}}{\max}\,{\rm Tr}\left[\frac{1+\mathbf{r}\cdot\boldsymbol{\sigma}}{2}\frac{1}{n}\sum_{j}A_{j}\mathbf{b}_{j}\cdot\boldsymbol{\sigma}\right]\nonumber\\
 & = & \underset{\mathbf{r}}{\max}\,{\rm Tr}\left[\frac{1}{2}\frac{1}{n}\sum_{j}A_{j}\mathbf{r}\cdot\mathbf{b}_{j}\right]\nonumber\\
 & = & \underset{\mathbf{r}}{\max}\left(\frac{1}{n}\sum_{j}A_{j}\mathbf{b}_{j}\right)\cdot\mathbf{r}\nonumber\\
 & = & \left|\frac{1}{n}\sum_{j}A_{j}\mathbf{b}_{j}\right|, 
\label{eq:parallelproof}
\end{eqnarray}
which is obtained only when $\mathbf{r}$ is parallel to $\sum_{j}A_{j}\mathbf{b}_{j}$. This proof also shows that the bound derived in Eq.~(\ref{eq:kbound}) is indeed a tight (attainable) bound.

Note that the maximal value of $S_n$ (the EPR-steering bound) is derived from a maximisation of both $\mathbf{r}$ and $\{A_j\}$, and the above maximisation, $\max(S_{n})|_{\{A_j\}}$, is only for the maximal value of $S_n$ that is attainable with a given $\{A_j\}$.
Conversely, it would be even simpler for a cheating Alice to optimise $\{A_j\}$ to obtain the maximal $S_n$ for a given $\rho_r$.
Knowing the orientation of Bob's state, Alice can easily deduce that for every measurement within $\frac{\pi}{2}$ radians of $\rho_r$, Bob's average result will be positive, and for every measurement more than $\frac{\pi}{2}$ radians from $\rho_r$, Bob's average result will be negative. To drive $S_n$ towards $-1$, Alice should choose $A_j=-1$ when $\langle \hat{\sigma}_{j}^{\beta}\rangle_{\rho_{\xi}^{\beta}}<0$, and $A_j=1$ when $\langle \hat{\sigma}_{j}^{\beta}\rangle_{\rho_{\xi}^{\beta}}>0$.
%

Clearly, the optimal values of $\{A_j\}$ corresponding to any $\rho_{\xi}^{\beta}$ are easily calculable, and the optimal orientation of $\rho_{\xi}^{\beta}$ is, almost as easily calculable for any $\{A_j\}$ permutation. However, $\rho_{\xi}^{\beta}$ can take any orientation on the Bloch sphere, whereas there are a finite number of  $\{A_j\}$ permutations. Therefore, a computational search for optimal cheating strategies is most efficient when calculating the optimal $\rho_{\xi}^{\beta}$ for every $\{A_j\}$, just as denoted in Eq.~(\ref{eq:kbound}).

Notably, when calculating the optimal cheating ensembles for each Platonic solid, it is found that the optimal orientations of $\rho_{\xi}^{\beta}$ are always either face-centred, or vertex-centred. This is intuitively understandable by observing the symmetries of the Platonic solids.

For the $n=3$ and $n=4$ solids, $n$ is equal to the number of corners on each face, and a point as close as possible to all corners of a shape will be equidistant from them. For $n=3$ and $n=4$, it is easy to see that a point equidistant from all $n$ vertices will be face-centred (the same logic holds for $n=2$, whose shape is a square, and its optimal ensembles being defined by the point in the middle of the line on any edge). However, this does not hold for the $n=6$ and $n=10$ solids, whose faces have $n/2$ corners, and thus for $n=6$ and $n=10$, there is no point that can be equidistant from all vertices. But intuitively, a point in the centre of $n$ vertices will still be an optimal orientation if those $n$ vertices are as close as possible. It turns out that the best arrangement of $n$ vertices for $n=6$ and $n=10$ will always be centred upon one of those vertices.

\begin{figure}
\begin{center}
\includegraphics[width=.9\linewidth]{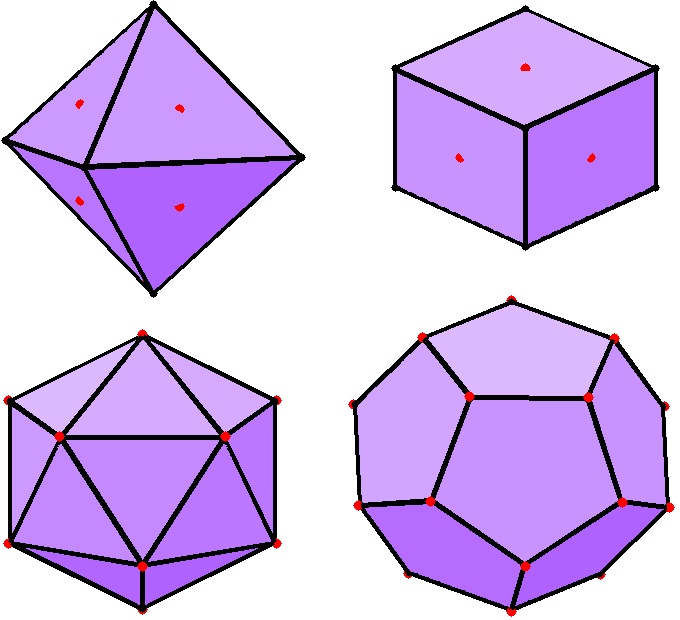}
\end{center}
\caption{(Colour online) Optimal cheating ensembles for the Platonic solid measurements, represented by the orientations of the red dots shown, relative to the vertices of each shape. From top left to bottom right, $n=3,4,6,10$ ($n=2$ not pictured).}
\label{fig:Efficient}
\end{figure}

The reader may have deduced that this reasoning does not lead to just one optimal ensemble for any measurement set. Indeed, as shown in Fig.~\ref{fig:Efficient}, there are multiple equally optimal cheating ensembles for each shape, their multiplicity equalling the number of the solids' faces for $n=3$ and $n=4$, and the number of the solids' vertices for $n=6$ and $n=10$.

However, the above ensembles do not strictly describe the optimal cheating strategies for $W_n$. The proof that the bounds $g_n$ on $W_n$ are also tight is somewhat different. The actual values of Alice's results (if choosing between $A_j=1$ and $A_j=-1$) are of no importance in any cheating strategy for maximising $W_n$, as long as Alice chooses $A_j$ deterministically for any given $\rho_{\xi}^{\beta}$. Recall that the bounds are defined as the maximal eigenvalues of $\frac{1}{n}\mathbf{N}$, and any eigenvectors that attain these eigenvalues are used to define $\mathbf{r}$, the optimal orientations of Bob's state. Since $\frac{1}{n}\mathbf{N}=\frac{1}{3}\mathbb{I}$ for the Platonic solids,  any vector is an eigenvector and hence any pure state is equally optimal for $W_n$.

The $S_n$ and $W_n$ functions we have discussed so far are those derived under the assumption that Alice submits results in every run of the experiment. We now turn to the issue of dealing with loss.

\section{Dealing with Loss}

In all of the above analysis, the only experimental imperfection we discussed was that the entangled state may not be pure. Typically a much greater concern, with photons at least, is inefficient detectors and other types of loss. The focus of this paper is to derive rigorous EPR-steering inequalities which are robust in the presence of the inevitable experimental detection inefficiencies and other losses. We discuss four ways of dealing with loss, which we identify as denial, anger, depression, and hope.

\subsection{Dealing with loss: Denial}

In most experiments involving photon polarisation, one has a very good knowledge of the system under investigation, and of the mechanism of the detector. This makes it possible to draw reasonable conclusions about which results are more likely to have been ``omitted'' when a photon is not detected. Indeed, by careful construction it is possible to ensure that the probability of omission is independent of what the result ``would have been.''

Thus, to draw conclusions about the behaviour of a system under study with imperfect detectors, it is common (and, \textit{a priori}, reasonable) to post-select one's results such that all null results are ignored, since null results are generally caused by properties of one's detectors, and not the system under study. Properly applied, this leads to post-selected ensemble measurements more accurately reflecting the properties of the initial ensemble. However, because omitted results, by definition, cannot be known, this conclusion cannot be proven. Thus, it is referred to as the fair sampling assumption (FSA). It is still a useful assumption, as it is based on accepted physical principles. For experiments that involve calculating probability distributions or averages (as very many experiments on quantum properties do), assuming that the omitted results obey a known probability distribution is often essential for obtaining meaningful results. This has been employed in a number of papers on tests of EPR-steering \cite{Saunders,Parsimonious}, where closure of the detection loophole was not a concern.

However, the FSA is an assumption based on the principles of quantum mechanics, and therefore should not be applied as a part of experiments intended to test the validity of quantum mechanics itself. Therefore, to apply this kind of post-selection to the results of an EPR-steering experiment would compromise its rigour. Similarly, if we are using EPR-steering to prove the existence of \textit{entanglement} when one party (Alice) is genuinely untrustworthy, then by using the FSA we are in \textit{denial} about the problem of loss, as we will now show.

Making the fair sampling assumption with a potentially dishonest Alice allows a cheating Alice to violate EPR-steering bounds that incorporate the FSA. This can be seen by reconsidering the expression for $S_{n}$, Eq.~(\ref{eq: Sn}),
\begin{equation}
S_{n}=-\frac{1}{n}\sum_{j}^{n}A_{j}\langle\hat{\sigma}_{j}^{\beta}\rangle_{\rho_{\xi}^{\beta}},\label{eq:Sn}
\end{equation}
and for its corresponding bound $k_{n}$, Eq.~(\ref{eq:kbound}),
\begin{equation}
k_{n}=\underset{\{A_{j}\}}{{\rm max}}\left[\lambda_{{\rm {\rm max}}}\left(\frac{1}{n}\sum_{j}A_{j}\hat{\sigma}_{j}^{\beta}\right)\right].
\end{equation}

If Alice claims not to have a perfect detector, she has the option of claiming that she did not receive a result from her detector on certain measurements. In these cases, Bob (who, for now, we assume has a perfect detector) will discard Alice's null results when calculating $S_{n}$ if he is making the FSA.
This post-selection of Alice's results rules out a rigorous test of EPR-steering. To illustrate this, consider a result of Eq.~(\ref{eq:Sn}) 
for any case in which $|\langle\hat{\sigma}_{j}^{\beta}\rangle_{\rho_{\xi}^{\beta}}|$ are 
different for (at least some) different values of $j$.
If Alice chooses to report null results for the lowest-valued $|\langle\hat{\sigma}_{j}^{\beta}\rangle_{\rho_{\xi}^{\beta}}|$, and Bob post-selects out these null results, $S_{n}$ will be higher than if Alice submitted results with perfect efficiency. The same goes for the cases of Alice submitting nulls for any number of measurements whose expectation values are less than the maximum $\langle\hat{\sigma}_{j}^{\beta}\rangle_{\rho_{\xi}^{\beta}}$ value. Indeed, if $|\langle\hat{\sigma}_{j}^{\beta}\rangle_{\rho_{\xi}^{\beta}}|$ equals unity for some $j$ (i.e., the state 
Alice sends is aligned with one of Bob's 
measurement axes $j$), then Alice could feign EPR-steering arbitrarily well, by omitting all those results except those for which 
Bob's measurement is aligned with $\rho_{\xi}^{\beta}$, when Bob is using a FSA. 
Thus, a FSA would be of benefit to an honest Alice, but would enable a dishonest Alice to cheat.
These comments apply equally to the nonlinear inequality [Eq.~(\ref{AVEPRS})] as well.


\subsection{Dealing with loss: Anger}

The inequalities derived above do not accommodate the possibility of Alice having an imperfect detector, and submitting any null results. One way to keep these inequalities is for Bob to require a result of Alice on every measurement -- Alice is not allowed to submit any nulls. If Alice is cheating, this will mean that her best option is to submit results according to her optimal strategies for perfect efficiency. If she tries to claim a null, and Bob demands a result nonetheless, a cheating Alice still has a foreknowledge of Bob's state that allows her to calculate an expectation value for $\langle {\hat{\sigma}^\beta_j} \rangle_{\rho^\beta_\xi}$, and make just as good an estimate of Bob's result (and thus, what result she should choose to maximise their correlations). On the other hand, if Alice is not cheating, she never has any knowledge of Bob's state except for that obtained from her own measurement result. So if she receives a null result, and Bob demands one of her anyway, she will have no way of estimating results that optimise the correlation function. Therefore, on average, she might as well choose $A_j = 1$ half of the time, and $A_j = -1$ half of the time. We imagine that being forced to make up random results would make an honest Alice angry, hence our name for this approach. If Bob chooses to do this, it will restore the rigour of the EPR-steering test, but it will mean that for an honest Alice with detector efficiency $\epsilon$, demonstration of EPR-steering will require $\epsilon\mu > k_n$ or $\epsilon^2\mu^2 > g_n$. This can be seen from separately calculating the contributions to $S_n$ and $W_n$ from when Alice does and does not receive a result, as we now show.

For an honest Alice with an efficiency of $\epsilon$, we can write 
\begin{eqnarray}
S_{n} & = & \left\langle -\frac{1}{n}\sum_{j}^{n}A_{j}\langle\hat{\sigma}_{j}^{\beta}\rangle_{A_{j}}\right\rangle _{A_{j}\in\{\pm1\} } \nonumber\\
& = & -\frac{1}{n}\sum_{j=1}^{n}\sum_{A_{j}=\pm 1}P(A_{j})A_{j}\langle\hat{\sigma}_{j}^{\beta}\rangle_{A_{j}}\nonumber\\
 & = & -\frac{1}{n}\sum_{j}^{n}\sum_{A_{j}}P(A_{j})A_{j}\left[\sum_{B_{j}}B_{j}P(B_{j}|A_{j})\right].
\label{AngerSn}
\end{eqnarray}
We can calculate the probability of Bob's results conditioned upon Alice's results by separating the cases where Alice does and does not register a detection. We distinguish these cases as $D=1$ and $D=0$, respectively, and use
\begin{equation*}
\sum_{B_{j}}B_{j}P(B_{j}|A_{j}) = \sum_{B_{j}}\sum_{D}B_{j}P(B_{j}|A_{j},D)P(D).
\end{equation*}

We can calculate the values of this expresssion, knowing that $P(B_{j}|A_{j},D=0)=\frac{1}{2}$ $\forall A_{j},B_{j}$; and that $P(D=1)=\epsilon$ and $P(D=0)=1-\epsilon$. From the nature of our entangled states, we can predict $P(A_{j}|D)=\frac{1}{2}$ for $D=1$, and as mentioned above, may as well be for $D=0$ also. We also know that $P(B_{j}|A_{j}=-B_{j},D=1)=\mu$, and $P(B_{j}|A_{j}=B_{j},D=1)=(1-\mu)$, and from these values, we can calculate that the value of $\langle\hat{\sigma}_{j}^{\beta}\rangle_{A_{j}}$ above is equal to $-A_j\epsilon\mu$. Since $P(A_{j}|D=1)=P(A_{j}|D=0)=\frac{1}{2}$ $\forall A_{j}\in[\pm1]$, we do not bother distinguishing between those cases in calculating the $P(A_{j})A_{j}$ prefactors, upon which we can find that $S_{n}=\epsilon\mu$ for our honest Alice. For the case of $\epsilon=1$, this agrees with the previously calculated result of Eq.~(\ref{WernerMu}).

By contrast, a dishonest Alice will always have some control over the value of $\langle\hat{\sigma}_{j}^{\beta}\rangle_{A_{j}}$, and in general can make it such that $\langle\hat{\sigma}_{j}^{\beta}\rangle_{A_{j}} \neq0$ whether she claims $D=1$ or $D=0$. Thus, she can submit $A_{j}$s such that $S_n$ always attains its maximum value of $k_n$. Therefore, to demonstrate EPR-steering under these conditions, an honest Alice would need $\epsilon$ and $\mu$ such that $\epsilon\mu>k_{n}$.

For the bounds on $W_n$, we repeat this working:
\begin{eqnarray}
W_{n} & = & \left\langle \frac{1}{n}\sum_{j}^{n}\langle\hat{\sigma}_{j}^{\beta}\rangle_{A_{j}}^{2}\right\rangle _{A_{j}\in\{\pm1\}}\nonumber\\
\label{AngerWn}
& =& \frac{1}{n}\sum_{j=1}^{n}\sum_{A_{j}=\pm 1}P(A_{j})\left[\sum_{B_{j}}B_{j}P(B_{j}|A_{j})\right]^{2}\\
 & = & \frac{1}{n}\sum_{j}^{n}\sum_{A_{j}}P(A_{j})\left[\sum_{B_{j}}\sum_{D}B_{j}P(B_{j}|A_{j},D)P(D)\right]^{2}.\nonumber
\end{eqnarray}
Using the statistics given above, the value of $\langle\hat{\sigma}_{j}^{\beta}\rangle_{A_{j}}$ here can be calculated to be $-A_j\epsilon\mu$, as above, and the value of $W_{n}$ obtained by an honest Alice under the restriction that $A_{j}\in\{\pm1\}$ will be $\epsilon^{2}\mu^{2}$. Again, for the same reasons as mentioned above, a cheating Alice will be able to attain the EPR-steering bound, $g_{n}$, on this criteria as well. Therefore, to demonstrate EPR-steering with this criteria (and the restriction that $A_{j}\in\{\pm1\}$), an honest Alice would need $\epsilon$ and $\mu$ such that $\epsilon^{2}\mu^{2}>g_{n}$.

\begin{figure}
\begin{center}
\includegraphics[width=.9\linewidth]{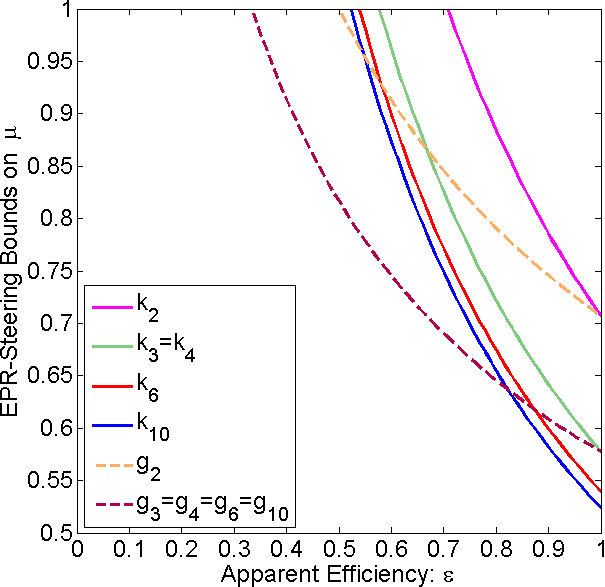}
\end{center}
\caption{(Colour online) The $k_n$ and $g_n$ EPR-steering bounds for the ``depression" strategies. The $k_n$ (linear inequality) bounds are the same for the ``anger" and ``depression" strategies, with $k_2,k_3=k_4,k_6,$ and $k_{10}$ corresponding, from right to left, to the solid lines. The $g_n$ (nonlinear inequality) bounds for anger and depression are easily distinguished, but $g_3=g_4=g_6=g_{10}$ in both cases. The dashed lines correspond to the depression bounds for $g_2$ (rightmost line) and $g_{n>2}$ (leftmost). The solid lines meeting the dashed lines at $\epsilon=1$ are also the anger bounds for $g_2$ (higher bound) and $g_{n>2}$ (lower bound): so $g_2=k_2$ and $g_{n>2}=k_{3,4}$ for anger.}
%
\label{fig:kndivep}
\end{figure}

Figure \ref{fig:kndivep} displays the values of $\mu$ required to violate the $k_n$ and $g_n$ bounds under this ``anger" strategy, as solid lines.
As mentioned in the previous section, the $g_n$ bounds are all equal for $n>2$, and this attribute remains for those bounds in Fig.~\ref{fig:kndivep}. One should also note that $g_n=k_3^2=k_4^2$ for $n\geq 3$
(and $k_2^2=g_2$ as well), and thus, the values of $\epsilon$ and $\mu$ required to violate the $g_2$ and $g_{n>2}$ bounds are, respectively, equal to those required to violate the $k_2$ and $k_{3}$ bounds. The nonlinear tests necessarily fail for $\epsilon < 1/\sqrt{3}\approx 0.577$. The linear tests necessarily fail for $\epsilon < 1/2$, 
because it is known that $k_\infty = 1/2$ \cite{key-3}.


\subsection{Dealing with loss: Depression}

An alternate approach which also maintains the rigour of the EPR-steering tests is to allow Alice to submit null results, but to incorporate these results into the EPR-steering function, as a zero value of any EPR-steering function that an honest Alice is contributing to (compared with Alice choosing $A_j = \pm1$ equally randomly, as with the ``anger" strategy above). 

However, even though an honest Alice can submit null results, because this strategy still does not entail any post-selection, her results would still have a reduced ($\epsilon$-dependent) correlation with Bob's results. We would expect an honest Alice to be depressed by this realisation, hence our name for this strategy. This reduction can be calculated just as above, with Eqs.~(\ref{AngerSn}) and (\ref{AngerWn}), by allowing Alice's null results to be represented by $A_j = 0$, for mathematical convenience (the additive inference variance criteria has previously been used with this approach for $n=2$ and $n=3$ \cite{Vienna,UQ}).

%
%

Note that this does not require any alterations to, or reformulations of our EPR-steering criteria, as their derivations hold for any values of $A_j$ satisfying $|A_j|\leq 1$.
Prior to now, we have restricted $A_j$ to $\{\pm1\}$ because a cheating Alice who does not want to get caught would do the same.

For $S_{n}$ in this case, an honest Alice has slightly different measurement statistics to those used above -- specifically, now that null results are an option, it is no longer the case that $P(A_{j}|D=1)=P(A_{j}|D=0)=\frac{1}{2}$ $\forall A_{j}$, since an honest Alice would choose $P(A_{j}=0|D=0)=1$, and therefore $P(A_{j}|D=0)=0$ $\forall A_{j}\in\{\pm1\}$. This difference means that the average value of $\langle\hat{\sigma}_{j}^{\beta}\rangle_{A_{j}}$ in this case will be $\mu$, rather than $\epsilon\mu$. However, the prefactor $P(A_{j})$ is no longer $P(A_{j})=\frac{1}{2}$ for $A_{j}=\pm1$, but is now $P(A_{j})=\frac{\epsilon}{2}$ for $A_{j}=\pm1$. From this, we regain an $\epsilon$ factor, resulting in the $S_{n}$ value for an honest Alice being $S_n =-\epsilon\mu$ again, just as for the ``anger" strategy. An Alice using the linear inequality would indeed be depressed by this result.

In calculating $W_{n}$, however, the ``depression" strategy actually gives a different, and better, correlation than the ``anger" strategy. We again calculate $\langle\hat{\sigma}_{j}^{\beta}\rangle_{A_{j}}$ to be $\mu$ instead of $\epsilon\mu$, but now our recalculation of Eq.~(\ref{AngerWn}) gives $W_{n}=\epsilon\mu^{2}$, with the $\epsilon$ factor coming from the $P(A_{j})$ prefactor, for an honest Alice.

Thus far, we have used $\epsilon$ to denote the detector efficiency of an honest Alice. A cheating Alice uses no detector, and therefore has no experimental inefficiency. However, she is still capable of submitting null results, and Bob has no way of telling whether such nulls (or any results) are genuine, except from calculating EPR-steering criteria. Thus, Bob would simply always regard Alice's ratio of non-nulls to nulls as her \textit{apparent efficiency}. From here on, this is what we will mean by $\epsilon$.

Since a cheating Alice does not determine her $A_j$ values experimentally, but rather chooses them strategically, the same should be true of her null results. As we have mentioned in Sec.~IV~A, if a cheating Alice submits null results for any measurements, she can do so such that her $A_j=0$ terms correspond to the lowest values of $|\langle\hat{\sigma}_{j}^{\beta}\rangle_{\rho_{\xi}^{\beta}}|$ predicted for Bob's measurements upon $\rho_{\xi}^{\beta}$. Thus, the $\langle\hat{\sigma}_{j}^{\beta}\rangle_{\rho_{\xi}^{\beta}}$ terms to which Alice assigns $A_j = 0$ could be selected such that their average is less than $k_n$ (and equivalently, $<\sqrt{g_n}$), meaning that the average of the $\langle\hat{\sigma}_{j}^{\beta}\rangle_{A_j=\pm 1}$ terms would be greater than $k_n$ (and their square would be greater than $g_n$). Thus, it would be easily possible for a cheating Alice to obtain $S_n > \epsilon k_n$ (and $W_n > \epsilon g_n$).

However, even in the presence of this kind of cheating strategy, since $k_n$ and $g_n$ represent optimal strategies for $\epsilon=1$, it can be deduced that a cheating Alice is still bound by $S_n \leq k_n$ (and $W_n \leq g_n$), because if it were possible to have a portion $\epsilon$ of the $B_j$ results average out to a value greater than $k_n/\epsilon$ (or $g_n/\epsilon$), then whatever strategy led to this result would work better at $\epsilon=1$ than the $k_n$ (or $g_n$) strategy does at $\epsilon=1$. Therefore, since $k_n$ (and $g_n$) is defined by the optimal strategy at $\epsilon=1$, it must be true that only the presence of EPR-steering could yield $S_{n}>k_n$ or $W_{n}>g_n$. Thus, without calculating more sophisticated loss-tolerant bounds upon $S_n$ or $W_n$, under this regime, an honest Alice would only convince Bob of EPR-steering if she possessed $\epsilon$ and $\mu$ such that $\epsilon\mu > k_n$ or $\epsilon\mu^2 > g_n$.

Thus, when Bob allows $A_j \in \{0,\pm 1\}$, the nonlinear inequalities become more powerful than when he does not (as in the ``anger" strategy), as shown in Fig.~\ref{fig:kndivep}. Moreover, for $\epsilon \lesssim 0.82$, the nonlinear tests for $n\geq3$ become more powerful than any of the linear tests we have constructed up to $n=10$. Nevertheless, all the tests still fail for $\epsilon<1/3$, as we can see from  Fig.~\ref{fig:kndivep}.

\subsection{Dealing with loss: Hope}

Clearly, it is important to take Alice's detector efficiency into account when performing tests of EPR-steering. Although dealing with null results is difficult for a Bob who wishes to be completely rigorous, having seen that more powerful bounds can be obtained by allowing null results would surely give him hope that he might calculate yet more powerful EPR-steering bounds by further taking $\epsilon$ into account. Indeed, we will see that lower EPR-steering bounds are possible for both the $S_n$ and $W_n$ functions, upon a more detailed analysis of how these bounds should be calculated when $\epsilon < 1$.

The basic idea, developed in Sec.~V, is that it is possible to calculate to exactly what extent a cheating Alice can use a simulated degree of loss to imitate quantum correlations, and from this obtain EPR-steering bounds as functions of both $\mu$ and $\epsilon$. Such bounds have been previously calculated and experimentally employed for the additive correlation bound $S_n$ \cite{Bennett}. In Sec.~V we present that calculation in detail and also calculate the best bounds using $W_n$, and compare these two types of EPR-steering criteria, for $n=2$, 3, 4, 6, and 10 measurements.

\section{Loss-Tolerant EPR-Steering Inequalities} 

As discussed above, reporting null results can be somewhat advantageous for a cheating Alice if she does so with knowledge of the state she sends to Bob, and the setting $j \in \{1,...,n\}$ he tells her to use in a given run. Say that Alice plans to declare non-null results only for $m$ of the $n$ possible settings Bob may communicate to her. Then her optimal strategy will be to send Bob a state $\rho_{\xi}^{\beta}$ that is better aligned with the $m$ measurements that Bob will make when Alice declares a non-null result. This strategy will be referred to as a ``deterministic strategy''. 

Note that if using only a single deterministic strategy, Alice could only feign efficiencies of $\epsilon=\frac{m}{n}\in\{\frac{1}{n},\frac{2}{n},...,\frac{n-1}{n},1\}$. However, Alice does not need to constrain herself to a single value of $m$ for the entire experiment (different runs may elicit different apparent efficiencies). If Alice uses multiple deterministic strategies -- a {}``nondeterministic strategy'' -- she is able to feign any measurable efficiency between $\epsilon =0$ and $\epsilon = 1$.

\subsection{Additive correlation bound}

The maximal bounds on $S_{n}$ obtainable by using deterministic strategies will be labelled $D_{n}(\epsilon)$. As we will see later on, some deterministic bounds on $S_{n}$ are not true bounds, because a dishonest Alice can do better with a nondeterministic strategy, even with $\epsilon = m/n$. The values of $D_{n}(\epsilon_m)$ are calculated from

\begin{equation}
D_{n}(\epsilon_m)=\underset{\{A_{j}\}_{m}}{{\rm max}}\left[\lambda_{{\rm {\rm max}}}\left(\frac{1}{n}\sum_{j}A_{j}\hat{\sigma}_{j}^{\beta}\right)\right].
\label{eq:KnDet}
\end{equation}
where $\epsilon_m=\frac{m}{n} \in\{\frac{1}{n},\frac{2}{n},...,\frac{n-1}{n},1\}$, and the sum is effectively only over $m$ members of $\{j\}$ because the set $\{A_{j}\}_m$ is subject to the condition that $A_{j}=0$ for $m$ elements, and $|A_j|=1$ for the remaining $n-m$ elements.


We can use $D_n(\epsilon_m)$ to calculate the nondeterministic bound on the non-post-selected correlation function, which is
\begin{equation}
K_{n}(\epsilon)=\underset{\{w_{m}\}}{{\rm max}}\left[\sum_{m=1}^{n}w_{m}D_{n}(\epsilon_{m})\right],\label{eq:Kn-FSA}
\end{equation}
where the sum over $m$ accounts for all possible optimal deterministic strategies that Alice could use. There is no benefit for Alice to use suboptimal deterministic strategies in any situation, so they are not considered, and each deterministic strategy can be indexed by $\epsilon_{m}\equiv\frac{m}{n}$. The term $w_{m}$ denotes the probability with which Alice chooses each deterministic strategy, so it is constrainted by $\sum_{m=1}^{n}w_{m}=1$ and $\sum_{m=1}^{n}w_{m}\epsilon_{m}=\epsilon$. From this form, it is clear to see that $K_{n}(\epsilon_{m})\geq D_{n}(\epsilon_{m})\forall\epsilon_{m}$.

When $S_{n}$ is calculated to incorporate Alice's null results in its averaging, thus avoiding any post-selection, its optimal bounds are described by Eq.~(\ref{eq:Kn-FSA}). However, such post-selection is still convenient for many experiments. To imbue our bounds with the ability to be easily compared with post-selected results (and also to display them as lower bounds on the necessary $\mu$, rather than on $\epsilon\mu$), the numerical bounds that we present shall be formatted as maximal bounds on $S_{n}$ when it incorporates post-selection upon Alice reporting a non-null result. This is easily accomplished, as such bounds are simply given by
\begin{equation}
k_{n}(\epsilon)=\frac{1}{\epsilon}K_{n}(\epsilon).
\end{equation}

Performing this post-selection is not making the fair sampling assumption, as these post-selected bounds still require Bob to keep track of Alice's apparent efficiency. That is, $k_{n}(\epsilon)$ is a completely rigorous upper bound on the post-selected correlation function $S_{n}$, that is obtainable in a no-steering model.

\begin{figure}
\begin{center}
\includegraphics[width=.9\linewidth]{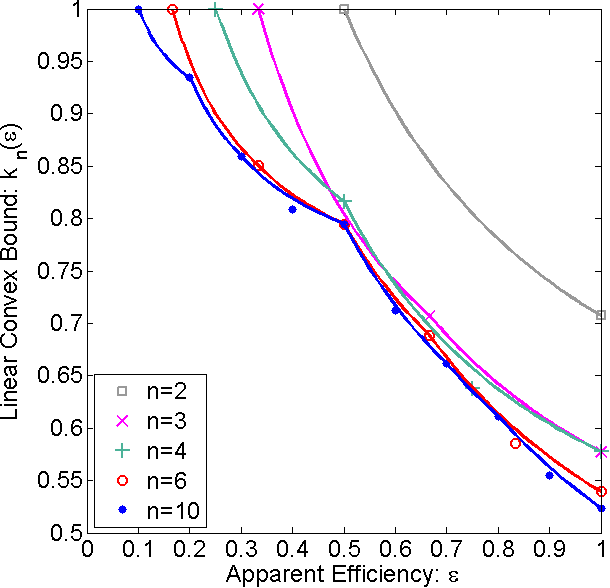}
\end{center}
\caption{(Colour Online) Deterministic (the points) and nondeterministic (the lines) post-selected bounds on $S_{n}$.}
\label{fig:CorBound}
\end{figure}

In a quantum mechanical model where Alice is honest, and she and Bob share a Werner state as in Eq.~(1), the calculated value of the post-selected $S_{n}$ will be $\mu$. This means that EPR-steering can only be demonstrated (by $S_{n}$) using states with values of $\mu>k_n(\epsilon)$, the lines in Fig.~\ref{fig:CorBound}. Note that for all $n$, EPR-steering cannot ever be demonstrated for $\epsilon\leq\frac{1}{n}$. With the loss-tolerant inequalities we derive here, that bound can be attained for $\mu \rightarrow 1$, unlike all previous approaches where the bound was $\epsilon = 1/3$.

If Bob uses detectors that possess some efficiency that is less than perfect, then it is important to note that he does not need to add any assumptions about what Alice is doing because the optimal LHS model (optimal for imitating EPR-steering) still cannot permit Alice to detect or control when Bob's detector does not yield a result. We trust that the FSA is valid for Bob's detector because we can trust that Bob's measurement efficiency cannot be affected by LHVs in any way that can assist a dishonest Alice in violating the EPR-steering inequality.

\subsection{Alice's inefficient cheating strategies for $S_n$}

In Fig.~\ref{fig:CorBound}, one should notice that in some places, the nondeterministic bounds are indeed greater than the deterministic bounds (and are never lower than the deterministic bounds). The deterministic bounds have been included (as markers) for the purpose of showing this.

\begin{figure}
\begin{center}
\includegraphics[width=.9\linewidth]{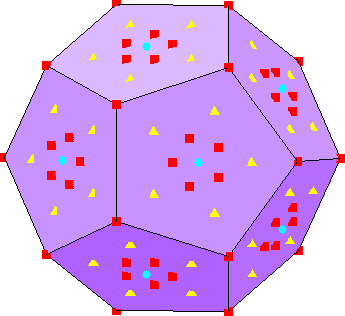}
\end{center}
\caption{(Colour online) Optimal spin orientations for $D_{10}(\epsilon)$ when $\epsilon = 0.3$ (yellow triangles), 0.4 (red squares), and 0.5 (blue dots).}
\label{fig:D10w345}
\end{figure}

For example, we can see that for $k_{10}(\epsilon=0.4)$, the nondeterministic bound does quite better than the deterministic bound. This is because a nondeterministic
combination of the deterministic $\epsilon=0.3$ and $\epsilon=0.5$ strategies performs better in this case. 
The three optimal deterministic strategies at $\epsilon = 0.3$, 0.4, and $0.5$ are shown in Fig.~\ref{fig:D10w345}. 
We remind the reader that the measurements available to Bob are defined by the vertices of Platonic solids, 
so it is easy to represent them graphically by plotting the solids that they describe (as in Fig.~\ref{fig:Efficient}). These points are connected by the edges defining the shape so as to make them more obvious. The other coloured points (or lines, later)
in the figure represent the spin orientations of the states that allow Alice to obtain the maximal deterministic bound (depending on her choice for $\{A_{j}\}$).

The ensemble orientations corresponding to the optimal deterministic strategies for $\epsilon=0.3$ are the points which are as close as possible to three vertices at a time. For $\epsilon=0.4$, the optimal deterministic choices are those which are close to four nearby vertices. Note that there are two classes of states which satisfy this requirement equally well - five on each face (similar to those for $\epsilon=0.3$) is one class, and one on each vertex is the second class.
For $\epsilon=0.5$, the optimal orientation must be close to five vertices, which corresponds to a face-centred orientation on the Platonic solid shown.

Upon close inspection of Fig.~\ref{fig:CorBound}, it 
is apparent that an equal mixture of the deterministic $\epsilon=0.3$ and $\epsilon=0.5$ strategies is better than the $\epsilon=0.4$ strategy. This can be explained as follows. The distance between any optimal spin orientation and its $m$ closest measurements is greater for $m=5$ than it is for $m=4$ (on average), and greater for $m=4$ than it is for $m=3$ (on average).
However, this separation increases more sharply from $m=3$ to $m=4$ than it does for $m=4$ to $m=5$. This difference in gradient is enough that the average separation (between measurement and optimal spin orientations) is greater for $m=4$ than it is for a weighted average of $m=3$ and $m=5$ strategies -- weighted such that every $\frac{3}{3+5}$ of the time, a $m=3$ strategy will be used, and every $\frac{5}{3+5}$ of the time, a $m=5$ strategy will be used. Using this weighting of terms, it can also easily be seen that the weighted average of $d_{10}(\epsilon=0.3)$ and $d_{10}(\epsilon=0.5)$ is greater than $d_{10}(\epsilon=0.4)$,
and indeed corresponds to the nondeterministic bound $k_{10}(\epsilon=0.4)$. This is the case for every deterministic strategy that is below the line of nondeterministic bounds. 

One may note from the above reasoning that our analysis of Alice's optimal cheating ensembles is relatively unchanged from that derived in Eq.~(\ref{eq:parallelproof}), except applied only to whichever measurements Alice chooses to be non-null. Indeed, this proof is still completely valid for determining Alice's optimal deterministic strategies regardless of whether or not the set $\{A_j\}$ includes any $A_j=0$ elements. Her optimal choice, given whatever measurements remain non-null for Bob, is still calculated in the same way: Her optimal orientation of the ensemble is simply a spatial average of the orientations of the remaining measurements.

\begin{figure}
\begin{center}
\includegraphics[width=.9\linewidth]{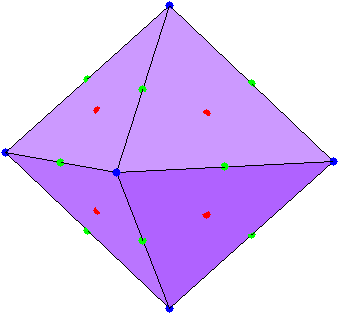}
\end{center}
\caption{(Colour online) Optimal spin orientations for $D_{3}(\epsilon)$ when $\epsilon = \frac{1}{3}$ (vertex-centred blue points), $\frac{2}{3}$ (edge-centred green points), and 1 (face-centred red points).}
\label{fig:n3mess}
\end{figure}

This is even more apparent for the simpler case of $n=3$ measurements, shown in Fig.~\ref{fig:n3mess}. Here, we can see the face-centred optimal ensembles, corresponding to the spatial averages over all trios of measurements, and yielding the highest values of $S_n$ when Alice submits no null results. Similarly, the edge-centred ensembles correspond to the closest possible points to any two vertices at a time, for when Alice omits one measurement out of every three. Finally, there are the vertex-centred points, which are optimal when Alice submits non-null results for only one measurement out of every three.

\begin{figure}
\begin{center}
\includegraphics[width=.9\linewidth]{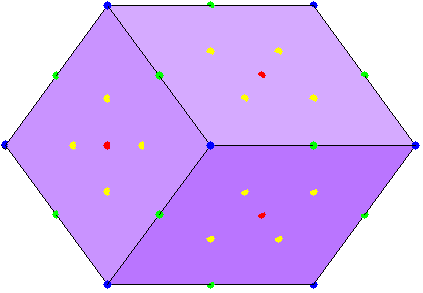}
\end{center}
\caption{(Colour online) Optimal spin orientations for $D_{4}(\epsilon)$ when $\epsilon = 0.25$ (vertex-centred blue points), 0.5 (edge-centred green points), 0.75 (yellow points between face-centres and vertices), and 1 (face-centred red points).}
\label{fig:n4mess}
\end{figure}

The optimal ensembles for $n=4$ measurements, shown in Fig.~\ref{fig:n4mess}, can be described in almost exactly the same way for the face-centred ($m=n$), edge-centred ($m=2$), and vertex-centred ($m=1$) sets of optimal ensembles. But there is the addition of the $\epsilon=0.75$ ($m=3$) ensembles here, which are the points as close as possible to three vertices as a time -- which is not as 
natural a task as it was for $n=3$, and this is reflected in Fig.~\ref{fig:CorBound}, where we see that these ensembles are actually inferior to a nondeterministic mixture of the $m=2$ and $m=4$ ensembles.

\begin{figure}
\begin{center}
\includegraphics[width=.9\linewidth]{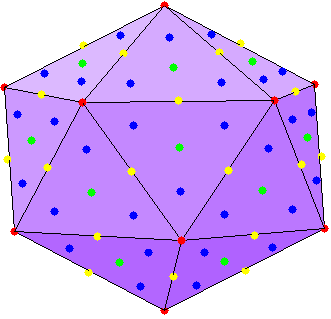}
\end{center}
\caption{(Colour online) Optimal spin orientations for $D_{6}(\epsilon)$ when $\epsilon = \frac{1}{6}$ (red vertex-centred), $\frac{2}{6}$ (yellow edge-centred), $\frac{1}{2}$ (green face-centred), $\frac{4}{6}$ (yellow edge-centred), $\frac{5}{6}$ (blue points between face-centres and vertices), and 1 (red vertex-centred). Note that because of the vertex symmetries, the optimal ensembles are identical for $D_{6}(\frac{1}{6})$ and $D_{6}(1)$, as is also the case for $D_{6}(\frac{2}{6})$ and $D_{6}(\frac{4}{6})$.}
\label{fig:n6mess}
\end{figure}

At a glance, the optimal ensembles for $n=6$ measurements, in Fig.~\ref{fig:n6mess}, are also quite comparable to $n=3$ in behaviour: the optimal ensembles for $m=1,2$ and 3, for obvious reasons. But the $\epsilon=1$ ensembles are not face-centred, but vertex-centred in this case, because the closest arrangement of six (non-antipodal) vertices on this shape will be centred on one of those vertices. Thus, the $m=6$ ensembles overlap with the $m=1$ ensembles. The $m=2$ and $m=4$ ensembles also overlap, because the closest arrangements of four vertices happen to be edge-centred. The $m=5$ ensembles, however, are more inelegant, and are over an edge's distance away from two vertices in each set of five that Alice submits non-null measurements for. It is not surprising that these are inferior to a nondeterministic mixing of the $m=4$ and $m=6$ strategies.

\begin{figure}
\begin{center}
\includegraphics[width=.9\linewidth]{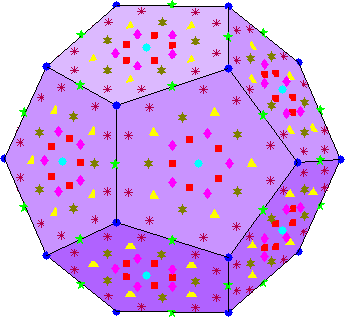}
\end{center}
\caption{(Colour online) Optimal spin orientations for $D_{10}(\epsilon)$ when $\epsilon = 0.1$ (dark blue dots), 0.2 (green five-pointed stars), 0.3 (yellow triangles), 0.4 (the red squares \textit{and} the dark blue dots), 0.5 (light blue dots), 0.6 (pink diamonds), 0.7 (brown six-pointed stars), 0.8 (green five-pointed stars), 0.9 (maroon asterisks), and 1 (dark blue dots).  Note that because of the vertex symmetries, the optimal ensembles are identical for $D_{10}(0.1)$ and $D_{10}(1)$, as is also the case for $D_{10}(0.2)$ and $D_{10}(0.8)$.}
\label{fig:n10mess}
\end{figure}

Figure \ref{fig:n10mess} displays the optimal ensembles for $n=10$ measurements, including those shown in Fig.~\ref{fig:D10w345}, for the purpose of illustrating the symmetries among the sets of arrangements. They will not all be discussed, but it is interesting to observe that there are overlaps between the $D_{n}(\frac{1}{n})$ and $D_{n}(1)$ strategies, as well as the $D_{n}(\frac{2}{n})$ and $D_{n}(n-\frac{2}{n})$ strategies, which was also the case for the $n=6$ ensembles (and is, put simply, a result of the similarities between the vertex-centred arrangements of the $n$ closest measurements on these two shapes). Note also the trend that optimal states move closer to face-centred until $\epsilon=0.5$, and then move back towards vertex-centred (which was also the case for $n=6$).

Something else of note is that for the Platonic solids -- because they are defined by regularly spaced vertices -- there is no particular measurement for which it is preferable for a cheating Alice to declare null results. When $n$ is high ($>4$), and $\epsilon$ is low, it becomes important to choose nulls such that the non-null measurements are close to one another, but there are no specific measurements that are ever \textit{more} optimal to remain null or non-null. 
However, even if some measurements were more preferable to omit, a smart cheating Alice would not do this, as it would lead to an unnatural pattern that a clever Bob would easily discover. Thus, for similar reasons as those in constraining $A_j = \pm 1$, it is strategically optimal for Alice to choose her sets of non-null results such that she submits $A_j=0$ equally often for each measurement (but offers no numerical advantage or disadvantage).

From numerical optimisation, and calculation of the $S_n$ and $W_n$ bounds, it can be seen that these same behaviours are evident in the optimal cheating ensembles for $W_n$, and indeed, the optimal deterministic strategies for $W_n(\epsilon)$ (which we will discuss below) were even observed to be the same strategies as above in some cases (though still quite different in others, as we shall see).

\subsection{Additive inference variance bound}

As shown earlier, the additive variance criterion is
\begin{equation}
W_{n}=\frac{1}{n}\sum_{j}^{n}E_{A_j}\left[\langle\hat{\sigma}_{j}^{\beta}\rangle^{2}_{A_j}\right].\label{eq:Wn}
\end{equation}
The function of $W_{n}$ does not explicitly depend on Alice's results, but implicitly uses them to calculate the expectation value of the above expression. The values of Alice's results are only
relevant for defining which of Bob's measurements contribute to this expectation value
and do not directly affect its outcome. Thus, for the case of a cheating Alice, the values that she defines for her non-null $A_j$ results will have no effect on the value of $W_n$, and therefore, there is no optimal strategy for choosing them. However, $W_n$ is affected just as $S_n$ by which results Alice chooses to be null, as Bob must still discard his measurements on those outcomes. 

The bound calculation for inefficient measurements is remarkably similar to that for when $\epsilon\equiv 1$ [refer to Eqs.~(\ref{gnworking}) and (\ref{AVEPRS})]. 
The only real difference is in the calculation of the matrix of Bob's measurements, $\mathbf{N}$, which we have reason to relabel as $\mathbf{N}_{m}$, seeing that
\begin{eqnarray*}
\sum_{j}\langle\hat{\sigma}_{j}\rangle^{2}_{A_j}
 & = & \mathbf{r}^{T}\left(\sum_{j}|A_j|\mathbf{b}_j\mathbf{b}_j^{T}\right)\mathbf{r}\\
 & = & \mathbf{r}^{T}\mathbf{N}_{m}\mathbf{r}\leq\lambda_{{\rm max}}(\mathbf{N}_{m}),
\end{eqnarray*}
where the included value of $A_j$ is not relevant to the maximisation [as it is in Eq.~(\ref{eq:KnDet})] except when its value is zero. Therefore, it has been included as $|A_j|$, which will only have any effect on this expression when $A_j = 0$ (the other values of $A_j$ being limited to $A_j=\pm1$).

However, taking care to treat $\mathbf{N}_{m}$ correctly, it can be seen that the deterministic bounds on $W_n$ are simply
\begin{eqnarray}
F_{n}(\epsilon_m)
 & = & \frac{1}{n}\underset{\{A_{j}\}_{m}}{{\rm max}}\left[\lambda_{{\rm max}}(\mathbf{N}_{m})\right]\nonumber \\
 & = & \underset{\{A_{j}\}_{m}}{{\rm max}}\left[\lambda_{{\rm max}}\left(\frac{1}{n}\sum_{j}|A_j|\mathbf{b}_j\mathbf{b}_j^{T}\right)\right]
\end{eqnarray}
where the maximisation over $A_j$ values, given $\epsilon_m$, is only over the choice of which are chosen to be zero or nonzero. The expression above has been arranged such that the manner of its similarity to Eq.~(\ref{eq:KnDet}) is most apparent. Indeed, this form of $F_n(\epsilon_m)$ can be used to calculate deterministic bounds as comparably as possible to $D_n(\epsilon_m)$.

Accordingly, the nondeterministic bounds on $W_n$ are calculated from weighted averages of $F_n(\epsilon_m)$, as
\begin{equation}
G_{n}(\epsilon)=\underset{\{w_{m}\}}{{\rm max}}\left[\sum_{m=1}^{n}w_{m}F_{n}(\epsilon_m)\right],
\end{equation}
with $w_m$ and $\epsilon_m$ being defined and used just as in the previous section's maximisation. 
This weighted averaging (and the constraints thereupon) directly corresponds to the way in which a cheating Alice composes her nondeterministic strategies. Thus, the maximal value of this expression (with respect to $\{w_m\}$) will correspond to the maximal value of $W_n$ that a cheating Alice is capable of attaining.
Therefore, $G_n(\epsilon)$ gives a tight bound on the post-selected 
 inference variance criteria. Again, we would wish to express this in terms of the maximal bounds on the non-post-selected inference variance criteria, 
\begin{equation}
g_{n}(\epsilon)=\frac{1}{\epsilon}G_{n}(\epsilon).
\end{equation}

\begin{figure}
\begin{center}
\includegraphics[width=.9\linewidth]{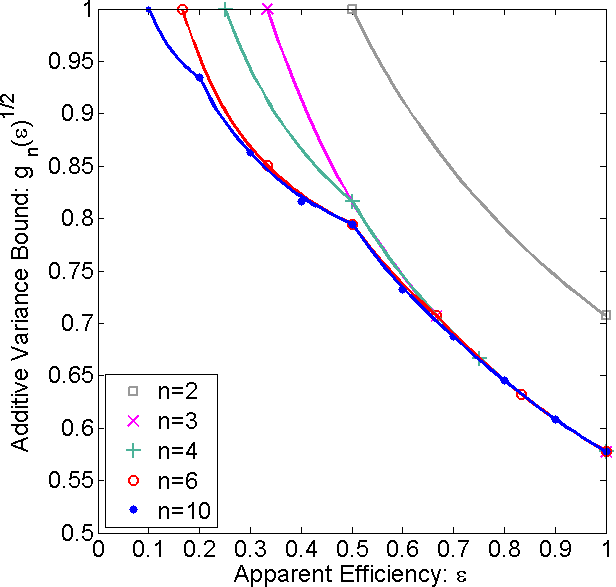}
\end{center}
\caption{(Colour online) Deterministic (the points) and nondeterministic (the lines) post-selected bounds on $\sqrt{W_{n}}$.}
\label{fig:VarBound}
\end{figure}

We could plot these bounds on $W_{n}$, but it will be more useful to plot the bounds on $\sqrt{W_{n}}$ because those values correspond to the bounds on $\mu$ required for a system to demonstrate EPR-steering. The bounds on $\sqrt{W_{n}}$ as a function of $\epsilon$ are given in Fig.~\ref{fig:VarBound}. The lines indicate the nondeterministic bounds, and the points represent the deterministic bounds. 
The shape of this graph is clearly quite similar to the bounds on $S_{n}$, aside from the fact that all of the bounds (except the $n=2$ bound) unite as $\epsilon\rightarrow1$. The bounds on $\mu$ given in the above graph are also similar to the bounds given by Fig.~\ref{fig:CorBound}, but a quantitative comparison will be given after the next subsection.


\subsection{Alice's inefficient cheating Strategies for $W_n$}


It is clear from Fig.~\ref{fig:VarBound} that the optimal bounds upon $W_n$ are indeed distinct from those upon $S_n$, but are still reasonably similar in their general behaviour. The 
measurement sets that yield the most similar results are those for $n=2,3,$ and 4. We can see -- and will show more clearly in the next subsection -- that both of the the deterministic bounds for $n=2$ (at $\epsilon=0.5$ and $\epsilon=1$) are unchanged from $S_n$ to $\sqrt{W_n}$. However, the nondeterministic mixtures of these two strategies do not lead to equal values of $S_n$ and $\sqrt{W_n}$, for the same reason that $x+y\neq\sqrt{x^2+y^2}$ in general [compare Eq.~(\ref{eq: Sn}) with Eq.~(\ref{eq:wn}) for clarification].
On the other hand, these expressions are equal when only one variable is nonzero, and from this comparision, we could infer that for any $n$-value, the $m=1$ strategies should always lead to the same bound on $\mu$ for $S_n$ or $W_n$. Indeed, this logic could also (correctly) lead us to expect the same deterministic strategies from either criteria in these cases, i.e., states aligned with each vertex on the relevant solids.

As discussed earlier, for any $n$-value's $\epsilon=1$ case, Bob's measurements form a spherical 2-design, and any LHS leads to the optimal bound. For the $n=2$ case, when $m=2$, this means that any state in the plane defined by these two measurements corresponds to an optimal deterministic strategy. Similarly, when $m=3$ for $n=3$, any state on the Bloch sphere is an optimal LHS.

However, for any $m=2$ case of the $n=3$ measurements, one might notice that whichever two measurements remain, they will take the same geometric form as the $\epsilon=1$ case of the $n=2$ measurements. Indeed, the calculation of $F_3(\epsilon=\frac{2}{3})$ for any two of these measurements is almost exactly the same as the working for $F_2(\epsilon=1)$ (with the same two measurements).

\begin{figure}
\begin{center}
\includegraphics[width=.9\linewidth]{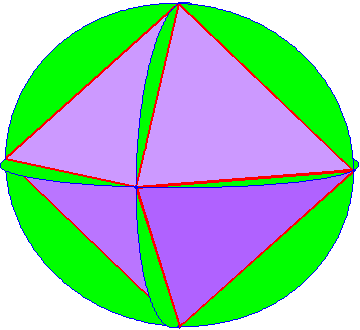}
\end{center}
\caption{(Colour online) Ranges of optimal spin orientations for $F_{3}(\epsilon = \frac{2}{3})$.}
\label{fig:n3range}
\end{figure}

The implication of this is that for $n=3$, the optimal deterministic strategies in any $m=2$ case will be the same as those for $F_2(\epsilon=1)$: That is, any states in the same plane as the two non-null measurements. Thus, Alice's optimal cheating strategies for $m=2$ are the planes shown in Fig.~\ref{fig:n3range}, with each plane being defined by a great circle on the Bloch sphere, and those planes being defined orthogonal to the orientation of the null measurement in each particular strategy.
So while the optimal deterministic bounds for $n=3$ are the same for $m=1,2,$ and 3 with $S_n$ or $W_n$, the optimal deterministic strategies for $m>1$ in either case are quite different (though the $S_n$ strategies are subsets of the $W_n$ strategies).

\begin{figure}
\begin{center}
\includegraphics[width=.9\linewidth]{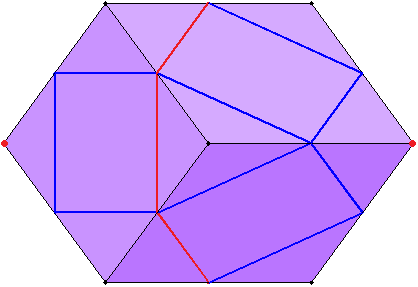}
\end{center}
\caption{(Colour online) Ranges of optimal spin orientations for $F_{4}(\epsilon = 0.75)$. The differently coloured (red) lines roughly down the centre of the figure correspond to the strategies that exclude the (red) vertex pair on the left and right tips of the figure above. Other (blue) lines correspond to other strategies. For $n=4$ and higher-$n$ shapes, this is a less cumbersome representation of planes than that used in Fig.~\ref{fig:n3range}.}
\label{fig:n4range}
\end{figure}

But Alice's optimal cheating strategies for $n=4$ are the same as in Fig.~\ref{fig:n4mess} for $m=1$ and 2 (and can be any LHS for $m=4$). However, the optimal LHS orientation for $m=3$ is not chosen by something so simple as the midpoint of two vertices, but is determined by a weighted average between three measurements. Thus, we have reason to expect that it will not be the same strategy for $S_n$ and $W_n$ here (and Fig.~\ref{fig:VarBound} shows a different deterministic bound at this point as well). Indeed, for $m=3$, the optimal $S_n$ strategies are not optimal $W_n$ strategies, and moreover there 
is actually a \textit{continuum} 
of optimal $W_n$ cheating strategies -- the range of optimal strategies is a discrete set of great circles  on the Bloch sphere, just as was observed with $n=3$ for $m=n-1$. To represent these strategies with maximal clarity, display of the circles has been foregone, and the lines on the cube's surface in Fig.~\ref{fig:n4range} are where the planes of these circles intersect the surface of the cube, and thus illustrate the positions of these ranges, relative to the cubic measurements.

The explanation for this is that when $m=n$, the matrix of Bob's measurements, $\mathbf{N}_m$, is equal to the identity matrix (multiplied by some constant), but for $m=n-1$, $\mathbf{N}_m$ is effectively the identity matrix minus one $\mathbf{b}_j\mathbf{b}_j^T$ projector. So this $\mathbf{N}_m$ matrix will still behave just like the identity matrix for any vector that is orthogonal to this $\mathbf{b}_j$ -- which is to say -- any vector orthogonal to $\mathbf{b}_j$ is an eigenvector of $\mathbf{N}_m$. 
The equivalence of the $\mathbf{N}_{m=n-1}$ matrix for $n=3$ with the $\mathbf{N}_{m=n}$ matrix for $n=2$ is a fortunate example for outlining how such a three-dimensional identity is simply reduced to a two-dimensional identity matrix in this way.
Thus, the spin orientations for each set of optimal deterministic cheating ensembles are all possible vectors in the plane orthogonal to the measurement that Alice chooses to be null in that particular strategy.

This is equally true regardless of which of the four projectors is the one omitted from $\mathbf{N}_m$, and each plane that is orthogonal to one measurement will also be orthogonal to its antipodal measurement. Therefore, there must be $n$ different planes of cheating ensembles, each corresponding to one of the $n=4$ different vertex pairs that represent the null measurement chosen in each respective cheating strategy. The lines on the cube that describe one of these planes have been differently coloured in Fig.~\ref{fig:n4range}, along with the pair of vertices that are orthogonal to this plane, and are thus the vertices Alice chooses to be null when using a LHS in this plane. The symmetries of these planes (visible from the number of different lines on each face of the cube) also reflect the number of different strategy sets being equal to $n=4$.

\begin{figure}
\begin{center}
\includegraphics[width=.9\linewidth]{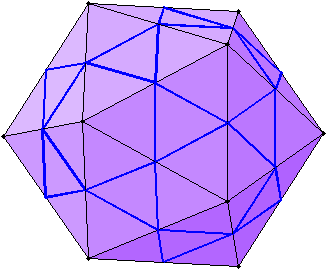}
\end{center}
\caption{(Colour online) Ranges of optimal spin orientations for $F_{6}(\epsilon = \frac{5}{6})$.}
\label{fig:n6range}
\end{figure}

\begin{figure}
\begin{center}
\includegraphics[width=.9\linewidth]{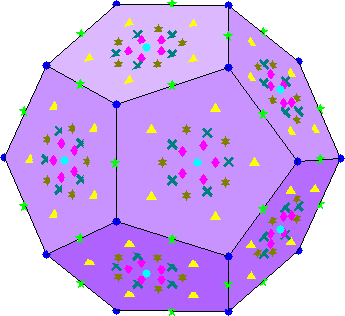}
\end{center}
\caption{(Colour online) Optimal spin orientations for $F_{10}(\epsilon)$ when $\epsilon = 0.1$ (dark blue dots), 0.2 (green five-pointed stars), 0.3 (yellow triangles), 0.4 (dark blue dots), 0.5 (light blue dots), 0.6 (pink diamonds), 0.7 (brown six-pointed stars), and 0.8 (aqua $\times$-marks). Note that because of the vertex symmetries' effect on $W_n$, the optimal ensembles are identical for $F_{10}(0.1)$ and $F_{10}(0.4)$, this time.}
\label{fig:n10messWn}
\end{figure}

This same behaviour occurs for the $n=6$ measurement set, where any set of deterministic cheating strategies using $m=5$ measurements, that is, $m=n-1$, will correspond to a plane of orientations orthogonal to the one null measurement. The arrangements of these planes relative to the $n=6$ measurement vertices is shown in Fig.~\ref{fig:n6range}. Because the angles between each face of the $n=6$ shape are more gentle than those of the $n=4$ shape, it is easier to see which lines compose individual planes, and which vertex pairs they are orthogonal to (so it was not necessary to highlight any strategies on this figure). 

Due to the highly symmetric arrangement of vertices for $n=6$, the optimal deterministic cheating strategies for $m=1,2,3,$ and 4 are still the same as those shown in Fig.~\ref{fig:n6mess}. This is understandable, looking at the orientations that are as close as possible to 2, 3, and 4 vertices at a time, which are all exactly in the centre of the closest arrangements of 2, 3, and 4 vertices. The same was not true of the $m=5$ strategies in Fig.~\ref{fig:n6mess}, so it is not surprising that those strategies are not included in the $m=5$ strategies for $W_n$, seen in Fig.~\ref{fig:n6range}.

\begin{figure}
\begin{center}
\includegraphics[width=.9\linewidth]{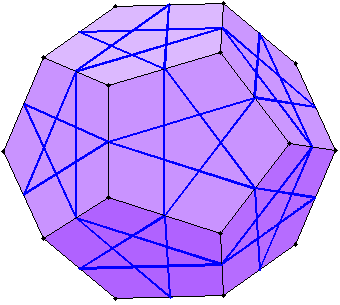}
\end{center}
\caption{(Colour online) Ranges of optimal spin orientations for $F_{10}(\epsilon = 0.9)$.}
\label{fig:n10range}
\end{figure}

In contrast with all of the above cases, when $n=10$ it is not just the $m=n-1$ deterministic ensembles 
that differ from the linear case. Indeed, this is true for half of the $n=10$ deterministic strategies in Fig.~\ref{fig:n10mess}, with the exceptions being the $m=1,2,5,8,$ and 10 strategies, which all fall either on a vertex, edge-centre, or face-centre. The optimal ensemble orientations for $W_n$ with $n=10$ measurements are shown in Fig.~\ref{fig:n10messWn}, excluding the optimal $m=9$ and $m=10$ strategies, neither of which are finite sets of points. As mentioned earlier, for $\epsilon=1$, any state is an optimal ensemble and, as expected, the optimal strategies for $m=n-1=9$ are the 10 planes which are orthogonal to the $n=10$ vertex pairs -- these are shown in Fig.~\ref{fig:n10range}.

\subsection{Comparison of bounds}

\begin{figure}
\begin{center}
\includegraphics[width=.9\linewidth]{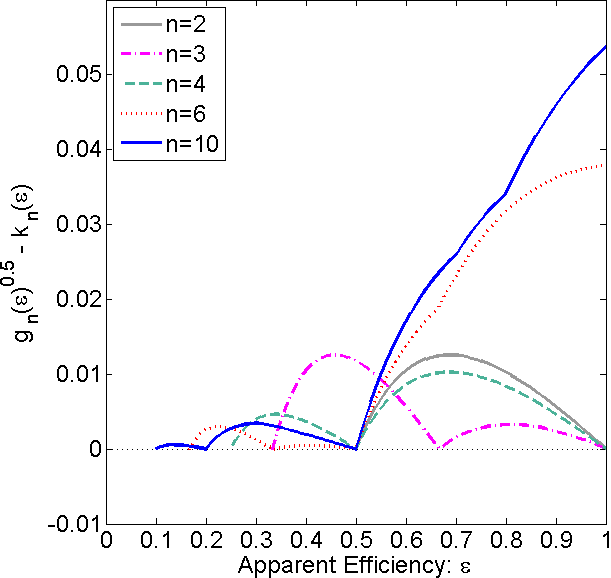}
\end{center}
\caption{(Colour online) The margin of improvement that the $S_n$ bounds have over the $W_n$ bounds for all $n$ and $\epsilon$ values (the absence of any negative values means that $k_n(\epsilon)$ is more loss-tolerant than $g_n(\epsilon)$ in all circumstances).}
\label{fig:Comparison}
\end{figure}

Because of the behavioural resemblances between the $k_{n}(\epsilon)$ and $\sqrt{g_{n}(\epsilon)}$ plots, plotting the data sets of Figs.~\ref{fig:CorBound} and \ref{fig:VarBound} on the same graph is not very illuminating since several lines are too closely related to distinguish. Indeed, upon close comparison, it can be seen that some of the $S_{n}$ and $\sqrt{W_{n}}$ bounds actually coincide (specifically, most of the deterministic bounds; see below). More significantly, it can also be seen upon closer inspection that $k_{n}(\epsilon)\leq\sqrt{g_{n}(\epsilon)}$ at every point. The optimal bounds on $\mu$ (out of those calculated in this paper) are thus the nondeterministic bounds in Fig.~\ref{fig:CorBound}, $k_{n}$. The advantage of the linear convex criteria over the additive variance criteria (for the $n=2$, 3, 4, 6 and 10 Platonic solid arrangements) is quantitatively displayed in Fig.~\ref{fig:Comparison}.

The discontinuous points in each line occur when either one (or both) of the criteria have a maximal bound that is also a nondegenerate deterministic bound. A nondegenerate deterministic bound may be defined as a deterministic bound that exceeds any other nondeterministic bound (since every deterministic bound is definable as a nondeterministic bound for which only one $w_m$ element is nonzero) for the same $\epsilon$. One may notice that the two bounds are equal at every such point except for those that occur for $n=6$ and $n=10$ when $\epsilon>0.5$.

\section{Conclusion}

The reality of loss (including inefficiency) is unavoidable in any rigorous test of EPR-steering, and in this paper we have explored several ways of dealing with loss.

In a rigorous test, Alice's null results cannot simply be accepted at face value and post-selected away, as the FSA is not a rigorous assumption, and one cannot deny (Sec.~IV~A) the loophole that its use would open. If Bob refuses to deal with detector losses, the only alternative is demanding Alice fabricate results in place of submitting nulls. This would anger (Sec.~IV~B) an honest Alice, who should be able to submit nulls as she receives them. For the criteria we consider, null results can still be submitted without being post-selected out if they are accepted as zero-valued results. Doing this in calculating the Additive Correlation criteria yields EPR-steering bounds that are, depressingly (Sec.~IV~C), unchanged. However, doing this for the Additive Variance criteria yields bounds that are somewhat improved. But the most hopeful (Sec.~IV~D) strategy for EPR-steering tests that are loss-tolerant is to allow post-selection of null results, while also calculating the highest possible criteria values that a cheating Alice can obtain with respect to her apparent efficiency.

This approach leads to the results displayed in Sec.~V, which are, for the two criteria we have explored, the lowest EPR-steering bounds obtainable when using the Platonic solid measurements. The two types of EPR-steering criteria presented in this paper are equally rigorous, but because it is less experimentally demanding, the additive correlation criterion would be considered the most useful of the two. The results shown in Fig.~\ref{fig:Comparison} are evidence of this. For convenience of comparison with existing experimental techniques and apparatus, the bounds plotted in this paper have been formatted to incorporate
post-selected results for Alice (so that they may be used to directly reference state entanglement parameter $\mu$ and Alice's detector efficiency $\epsilon$).

As we have seen from the results herein, rigorous experimental demonstration of EPR-steering cannot be performed using detectors of just any efficiency, or even for detectors of perfect efficiency, if one does not also possess ensembles of states that are sufficiently entangled. 
%
Even with the best test considered here ($n=10$), the purity of entanglement in one's states ($\mu$) must exceed $k_{10}(\epsilon=1)\approx0.52$, and then only with $\epsilon = 1$. In fact, it has been shown \cite{key-3} that even with an infinite number of measurements, EPR-steering is still impossible to demonstrate with any $\mu<0.5$, regardless of detection efficiency, but is demonstrable with any nonzero $\epsilon$ if $\mu$ is sufficiently high. This result is congruent with the pattern that we have seen, that $\mbox{\ensuremath{\epsilon}}=\frac{1}{n}$ is the lower bound on efficiencies for which EPR-steering is still demonstrable. If one wishes to forego measurements that are regularly spaced, then the choice of $n$ would not even be limited to $n \leq 10$, and a lower bound of $\mbox{\ensuremath{\epsilon}}=\frac{1}{n}$ would allow arbitrarily low $\epsilon$ values to yield successful EPR-steering tests, given arbitrarily large $n$.

The measurements used to derive the numerical results in this paper are oriented along the vertices of Platonic solids (with the exception of $n=2$), for the purposes of using measurements that are regularly spaced. The reason for this is that it should be most difficult for a LHS to imitate entanglement when the measurements performed on it are as different as possible. Thus, the Platonic solids are a good guess at Bob's optimal measurement strategies for $n=2,3,4,6$ and 10. However, this scheme offers no such measurements for any other $n$-value, and as mentioned in the conclusion, there would be clear benefits to using $n>10$ in some situations. In addition to this, there is no actual proof that the measurements used here are all optimal for $n=2,3,4,6$ and 10 measurements. Indeed, the contrary is easily proven by observing the overlap of $n=3$ and $n=4$ bounds around $\epsilon=0.5$ in Fig.~\ref{fig:CorBound} -- quite clearly, the $n=3$ bound performs better than the $n=4$ bound in this region, which could not happen if the $n=4$ measurements were optimal ones. These issues will be addressed in future work \cite{preparation}.


\section*{Acknowledgements}

This work was supported by the ARC Centre of Excellence CE110001027 and E.G.C. received support from Australian Research Council grants DP0984863 and DE120100559.

\end{document}